\documentclass[]{article}

\usepackage[letterpaper]{geometry}
\usepackage[table]{xcolor}
\usepackage{graphicx}
\usepackage{fullpage}
\usepackage[colorlinks,citecolor=black,linkcolor=black,urlcolor=black]{hyperref}
\hypersetup{colorlinks}

\title{Radial distribution gain at 633 nm in a He-Ne RF-excited small bore discharge}

\author{$\mathrm{Umberto~Giacomelli^{1,2,3,*}}$, Nicol\`o~$\mathrm{Beverini^{2}}$, $\mathrm{Angela~Di~Virgilio^{3}}$,\\ $\mathrm{Enrico~Maccioni^{2,3}}$ and $\mathrm{Paolo~Marsili^{2,3}}$}

\date{$ ^1$ GSSI, L'Aquila, Italy\\ $ ^2$ Dipartimento di Fisica, Universit\`a di Pisa, Pisa, Italy\\ $ ^3$ INFN sezione di Pisa, Pisa, Italy\\ $ ^*$ umberto.giacomelli@gssi.it }
\begin{document}

\maketitle

\begin{abstract}
Devices as Large Ring Laser Gyroscopes (RLG) for fundamental physics and
geophysics investigation are currently run by means of RF power supply systems.
This is not the standard method to supply a gas laser, that typically is powered with a DC system.
In literature RF power supply lasers have been studied several years ago, and to correctly understand the behaviour of
devices as RLG a more detailed study has been pursued.\\  
Detailed study of the radial distribution of the optical gain of an He-Ne discharge cell in function of gas pressure and RF power supply will be illustrated, discussed and compared with existing literature.
The presented analysis demonstrates that it is possible to optimize the RLG operation with a proper choice of gas pressure and power level of the RF power supply.
Accordingly we have been able to establish transversal and longitudinal single-mode operation our prototype GP2.
\end{abstract}

\section{Introduction}
Ring Laser Gyroscopes (RLG) are devices, based on the Sagnac effect, that measure with high sensitivity the rotational motion of the apparatus with respect to an inertia reference system.
By beating the two beams counterpropagating inside the ring optical cavity, an interference pattern at a specific frequency, the Sagnac frequency \cite{Schreiber2013} is produced, which is proportional to the rotation rate of the gyroscopes reference frame.
The ideal laser transition for this kind of application is the He-Ne 633 nm transition.
Small dimension devices, characterized by a ring optical cavity length, of the order of 40-50 cm, have been widely developed for application as high sensitivity navigation instruments.
The sensitivity can be increased by some orders of magnitude by larger frame apparatus, with cavity length ranging from some meters to tens of meters \cite{Schreiber2013,Gebauer2020a,Capozziello2021}, allowing applications to seismology, geodesy, and tests of fundamental physics.\\
Our research team is presently operating on two large frame ring lasers (LF RLGs), the first one, named GP2  \cite{Belfi2014,Santagata2015,Beverini2019a} installed in the INFN (Istituto Nazionale di Fisica Nucleare) laboratory in Pisa, and the second one, GINGERINO \cite{Belfi2016,Belfi2017,Belfi:18,DiVirgilio2020a,Capozziello2021} installed in the underground laboratory of Gran Sasso LNGS.\\
In any He-Ne gas laser, the gain is obtained by exciting a plasma discharge in the medium. 
This can be produced by a high voltage DC or by a Radio Frequency (RF) excitation.
In view of this, this work aims to study the optical gain of the line at 633 nm in a Radio-Frequency (RF) excited He-Ne plasma gas in function of both the radial behavior along the discharge bore and the total filling gas pressure.
The studies about this topic date back to several years ago \cite{Mielenz1965,Mazanko1971,Tsarkov1973,Spoor1984,Andrews1989,Gray1996}.
During years for commercial devices, the DC supply has been preferred over the RF one.
This is because it is easier to realize, and it avoids the generation of RF-noise.
However, DC discharge presents problems if high accuracy is required, due to the asymmetry introduced by the cataphoretic motion.
Due to this, all the LF RLGs use rf plasma excitation.
Moreover, it was shown a better efficiency of rf excitation in terms of optical gain in function of working power \cite{Gray1996}.\\
For application to LF gyros it is also important to extend the study to low gain and high buffer pressure conditions.
Indeed, a LF RLG must operate in a different regime than the small devices.
For short optical cavity length, the free spectral range (FSR) is large compared to the Doppler linewidth of the laser transition, so that the laser can easily operate stably on a single longitudinal mode.
This is no more the case in LF RLG.
However, it was demonstrated that also in those conditions stable single-mode operation can be obtained operating at low emission power, near laser threshold \cite{Bilger1995}.
Moreover, a higher value of the helium pressure can help to stabilize the single-mode operation by increasing the homogeneous linewidth. 
\section{Experimental setup}
\begin{figure}
	\centering
	\includegraphics[width=\textwidth]{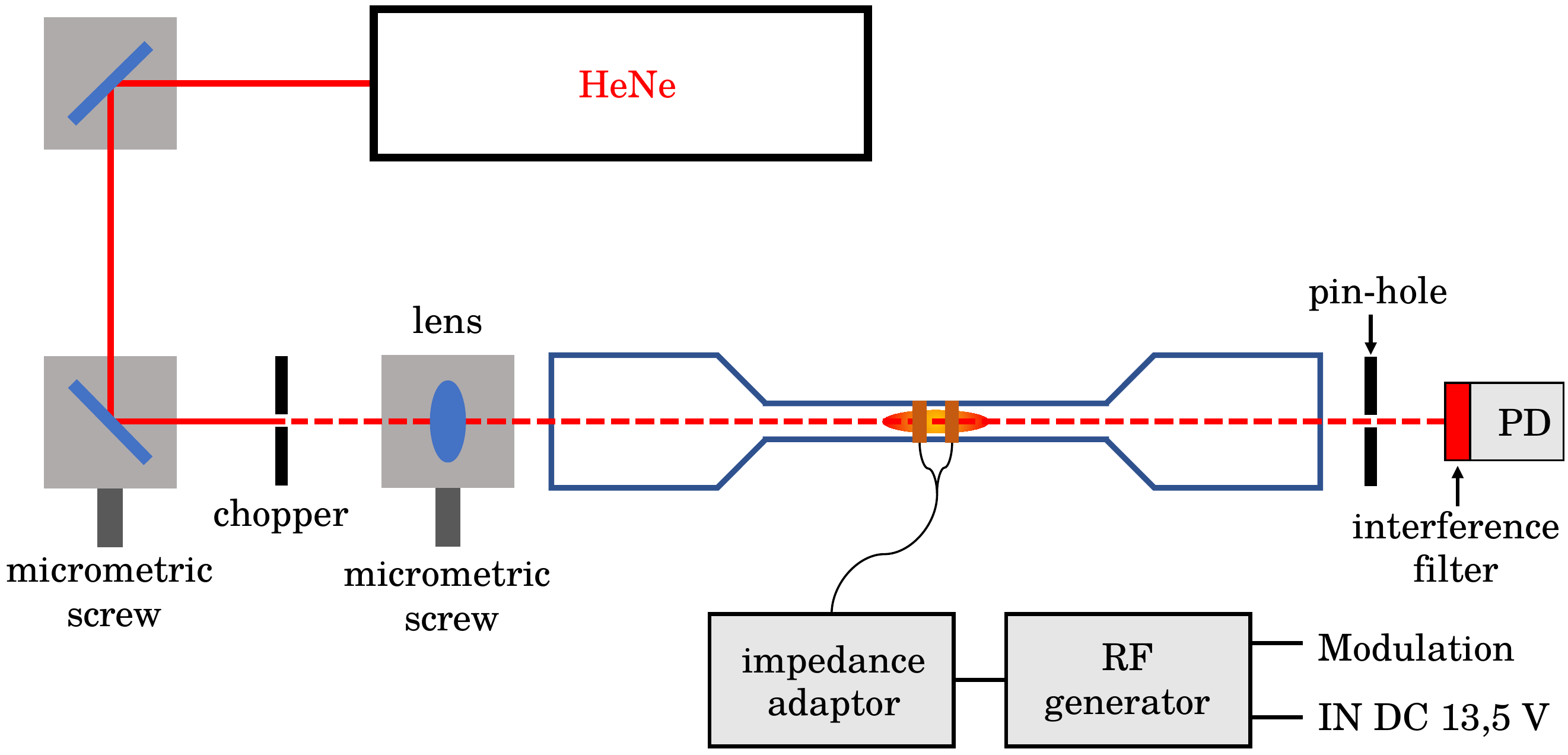}
	\caption{Optical Setup \label{fig_optical}}
\end{figure}

The opto-electronic setup is shown in Fig \ref{fig_optical}.
A He-Ne power stabilized laser is used as a probe to investigate the plasma discharge optical gain along its bore diameter.
The laser light is sent to the plasma discharge via two mirrors.
In order to move it through the plasma diameter the holder of the second mirror is mounted on a traslator with a micrometric screw.
Before the injection into the cavity there is a chopper and a plano-convex lens of 300 mm focal length.
The chopper modulates the intensity of the laser, and the lens focuses it in the middle of the plasma discharge.
In this way the waist has a size of $150~\rm \mu m$ in the center of the glass bore, which has a diameter of 5.6 mm and a length of $\rm 150~mm$.
Similarly to the second mirror, also the lens holder has a micrometric screw translation slide to preserve the alignment of the probe beam when it is moved along the plasma diameter.
In the middle of the tube there are two copper rings acting as electrodes for the plasma discharge.
Each one has a width of $\rm 10~mm$ and the distance between them is about $\rm 3~mm$, that we have identified as optimal distance to have an uniform discharge in a large range of RF power values.
The power supply is a radio frequency generator operating around 150 MHz\cite{Graham2006}.
A passive impedance adapter, built with a variable capacitor, links the electrodes with the RF generator and it is fundamental to minimize the reflected power and optimize the pumping efficiency.
We have realized a control circuit for the RF Amplifier Module in order to modulate its output power via a trimmer and via an input voltage channel.
At the end of the glass cavity there are a pin-hole and a photodiode with a narrow pass-band filter centered at 633nm.
The pin-hole, of a 1 mm of diameter, and the filter before the photodiode, shield it from the fluorescence emission of the plasma discharge.
The position of both, the pin-hole and the photodiode, are readapted to optimize the signal before each measurement.
\section{Measurement}
\subsection{Measurement technique}
We chose to investigate four different pressure of gas mixture and, for each pressure, three different power level of plasma discharge.\\
Following the literature \cite{Spoor1984}, we have chosen for each He-Ne mixture, a pressure of 0.22 mbar of Ne and then we have added He up to 1.33, 2.66, 5.00, 10.00 mbar.
Before the first filling the cavity has been baked and pumped down to $10^{-6}$ mbar, and before each refill we have restored the vacuum condition to clean the system.
Considering the size of the tube and the laser waist dimension, we chose to move the probe laser along the diameter of the plasma discharge by steps of 0.25 mm using the micrometric screws shown in Fig \ref{fig_optical} .\\
The difficulties of this type of study are mainly two: the handling of a RF source, and the fact that the expected gain of this kind of system is of the order of only some hundreds of ppm \cite{Mazanko1971,Tsarkov1973,Spoor1984,Gray1996}.\\
About the first issue, differently from a DC plasma discharge the power supplied by a RF source needs an initial high-voltage discharge to switch on the plasma. 
This peculiarity makes impossible to use the typical measurement procedure of a DC discharge that involves to turn off and on the plasma, to empathize its gain effect, during the measurement.
We solved this problem modulating the power of the plasma between two levels.
This method allows to evaluate the gain difference between this two plasma power levels.
It is possible to obtain the absolute value of the gain by properly choosing the value of the low level.
Setting this level near the plasma threshold, it is possible to have a stable plasma discharge with enough low power to not activate the gain for the 633 nm transition.\\
The second problem concerns the low expected gain around few hundreds of ppm at most \cite{Mazanko1971,Tsarkov1973,Spoor1984,Gray1996}.
In addition, the spontaneous emission at 633 nm wavelength makes it hard to discern between the laser gain and the offset due to the spontaneous emission.
Lock-in amplifier has been used to improve the measurement precision.
The chopper in Fig \ref{fig_optical} modulates the light amplitude coming into the cavity, and the Lock-In demodulates and amplifies the signal detected by the photodiode.
This method avoid any other effects except for the ones on the modulated light.\\
To have a statistically significant set of data, each measurements lasts 15 minutes, during which we modulate the plasma discharge between low and high levels with a square wave of 1s period. 
\subsection{Performed Measurement}
\begin{figure}
	\centering
	\includegraphics[width=\textwidth]{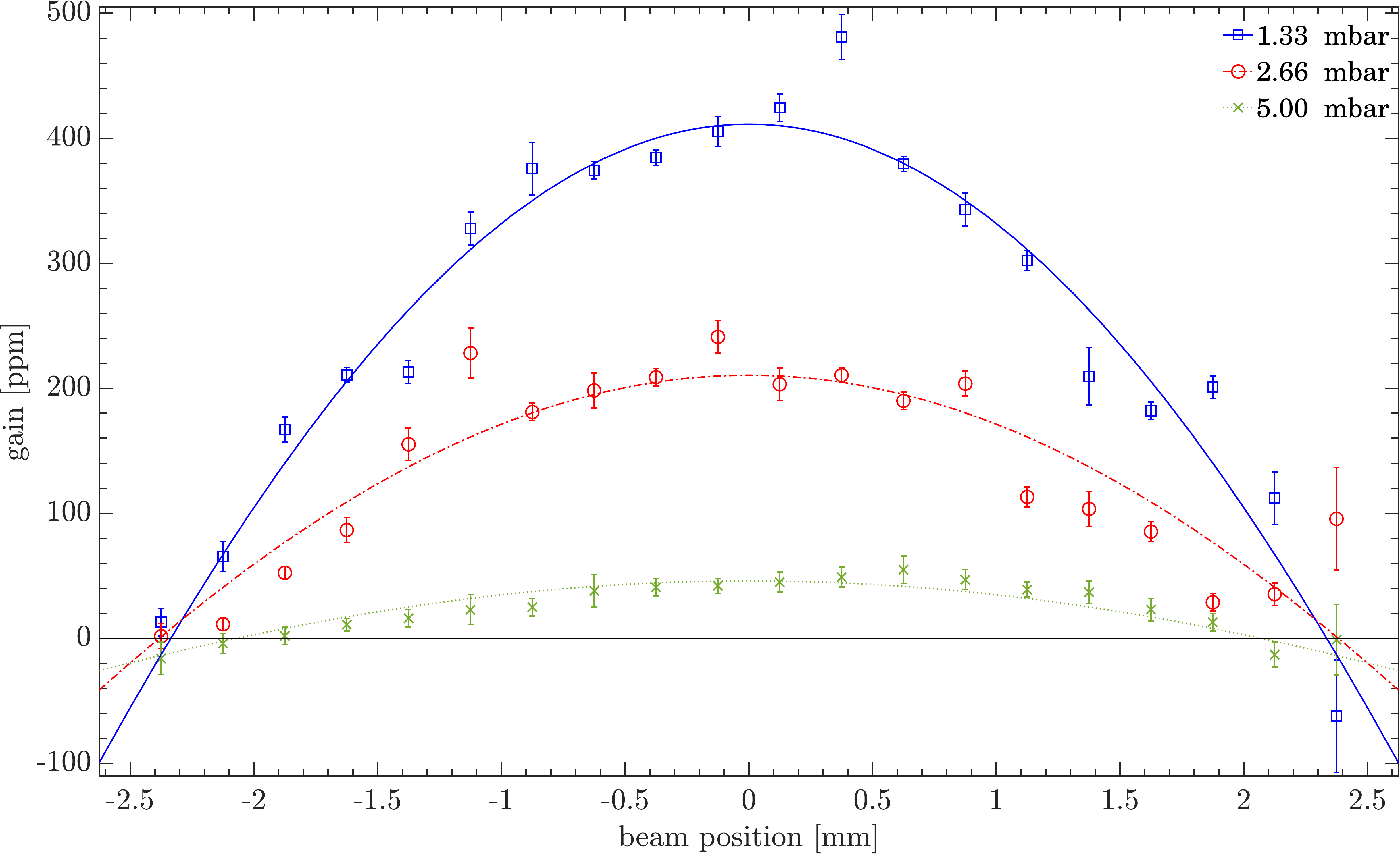}
	\caption{Optical gain vs beam position at RF generator output amplitude between 0.15 W (low level) and 0.4 W (high level) \label{fig_gain_1}}
\end{figure}

\begin{figure}
	\centering
	\includegraphics[width=\textwidth]{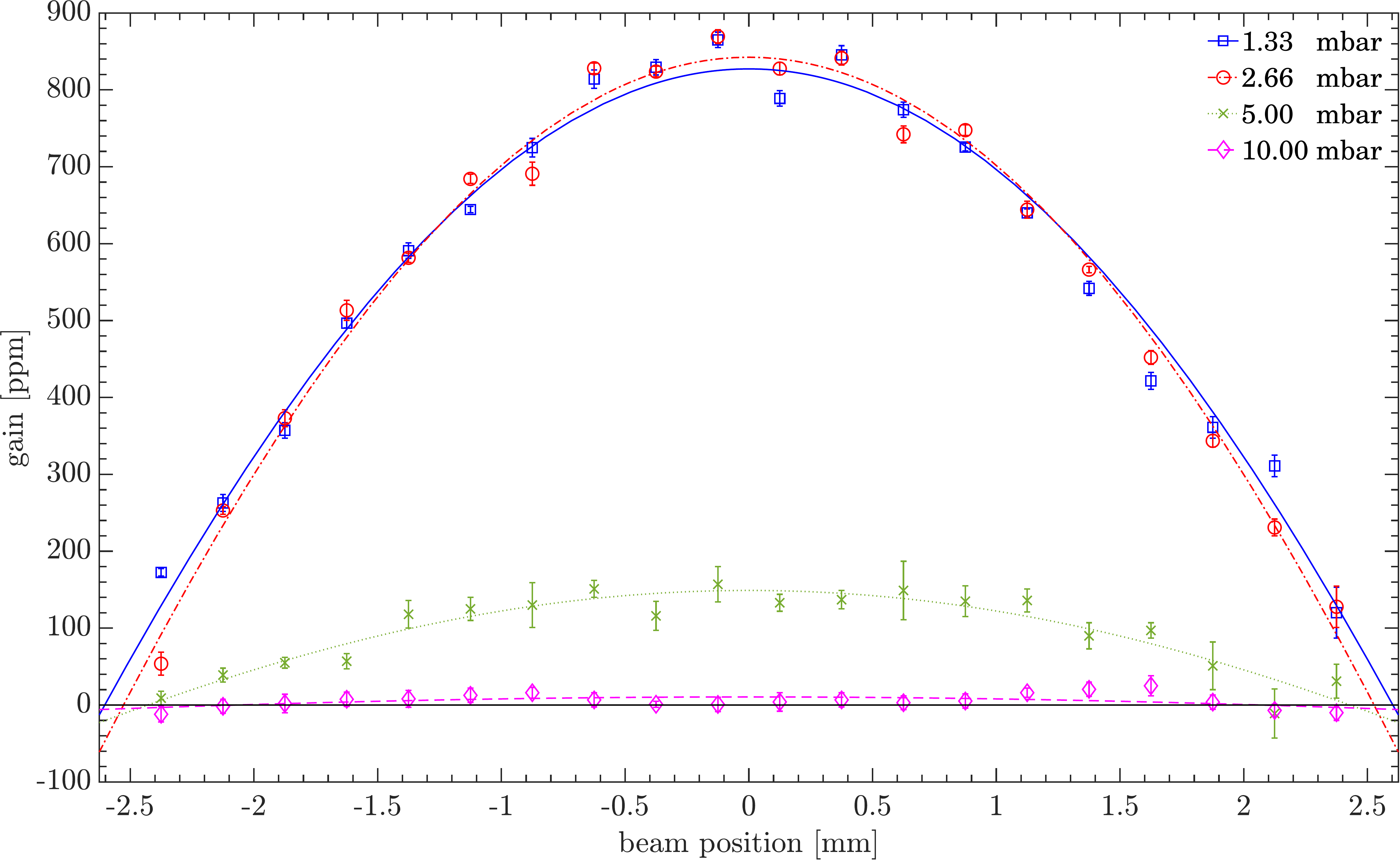}
	\caption{Optical gain vs beam position at RF generator output amplitude between 0.27 W (low level) and 1 W (high level) \label{fig_gain_2}}
\end{figure}

\begin{figure}
	\centering
	\includegraphics[width=\textwidth]{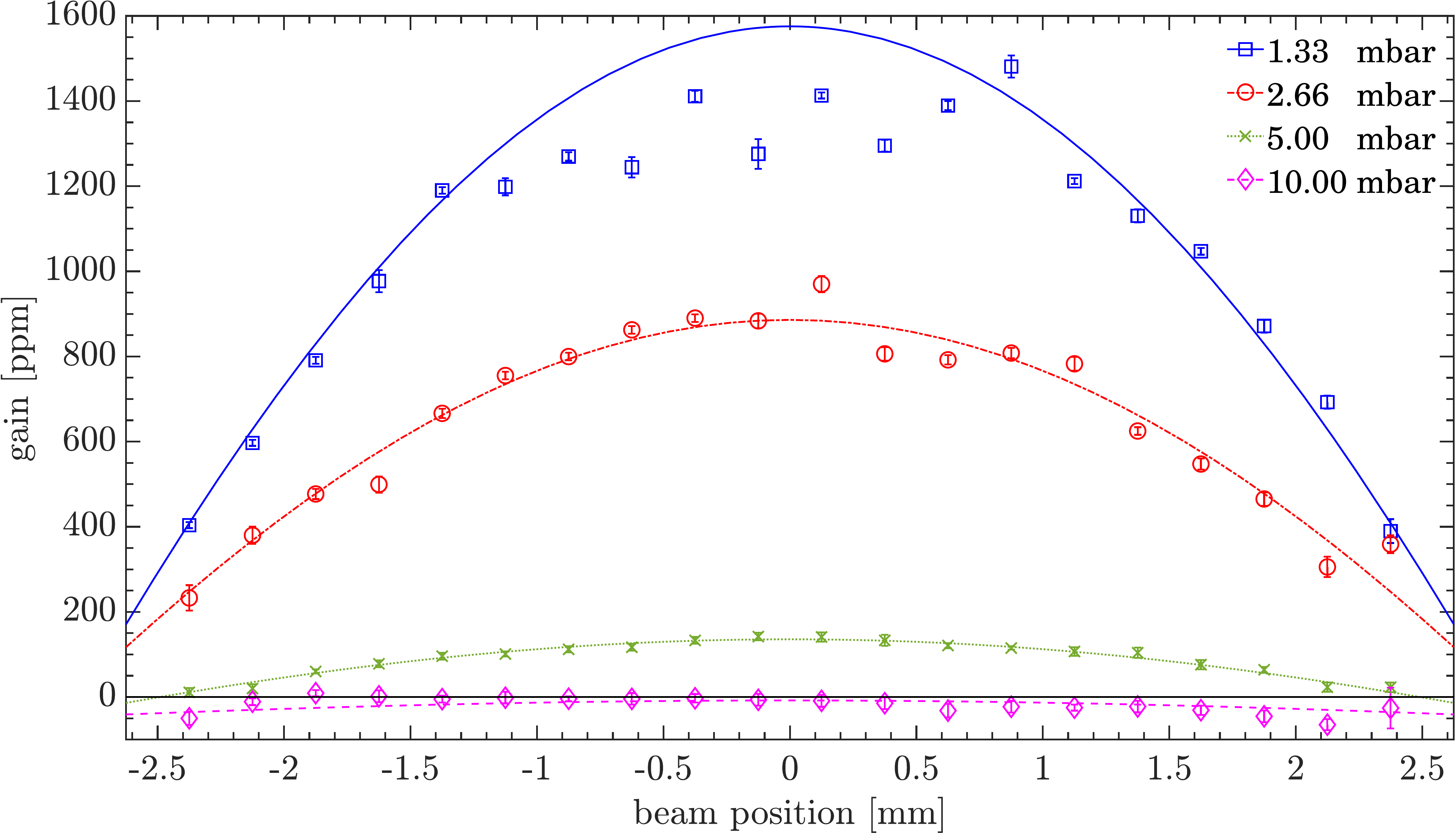}
	\caption{Optical gain vs beam position at RF generator output amplitude 0.22W (low level) and 2.2 W (high level) \label{fig_gain_3}}
\end{figure}

Figs \ref{fig_gain_1},\ref{fig_gain_2},\ref{fig_gain_3} show the gain data in function of the position along the bore diameter.
The continuous lines are the fit with the function shown in Eq. \ref{eq:McLeod_model}
\begin{equation}
	G(r)=a J_0\left(2.405 \frac{r}{R}\right) + b
	\label{eq:McLeod_model}
\end{equation}
where $a$ and $b$ are the fit parameters, $J_0$ is the 0 order Bessel function of the first kind, $R$ is the radius of the glass tube, and $r$ is the distance from the center at wich we evaluate the gain $G(r)$.
This is an approximation of the Mc Leod model \cite{McLeod2000a} used by R. Graham in his PhD Thesis \cite{Graham2006} for the electron population of the plasma.
This model considers the gain factor directly proportional to the electrons distribution, and how it is shown in the plots this hypothesis is in a good accordance with the experimental data.
Figures \ref{fig_gain_1}, \ref{fig_gain_2} and \ref{fig_gain_3} contain the data referred to a specific couple of low and high output power of the RF discharge source.
It is important to underline that, even if the low levels of the three plots are different we have verified that in any case we are enough close to the threshold level to not activate the 633nm transition in the plasma medium.
There are no data at 10.00 mbar in Fig \ref{fig_gain_1} because the low level of 0.15 W is under the threshold level for this gas pressure.
\\
To fully scan the glass tube diameter, each set of data is composed by 20 steps of the micrometric screw.
Considering the duration of each single measurement and the dead time between successive steps, the total time needed for each set of data is about 6 hours.
During this time the hydrogen, released by the glass bore, contaminates the gas mixture.
The small volume of gas and the absence of a getter pump makes this contamination not negligible in the evaluation of the plasma gain effect.
To correct this behavior we have extracted the time dependency of the gain.
To do that, we have acquired the gain with the same method described before, but without moving the laser probe through the diameter.
This allows to obtain the gain behavior in function of the time.                   
\begin{figure}
	\centering
	\includegraphics[width=\textwidth]{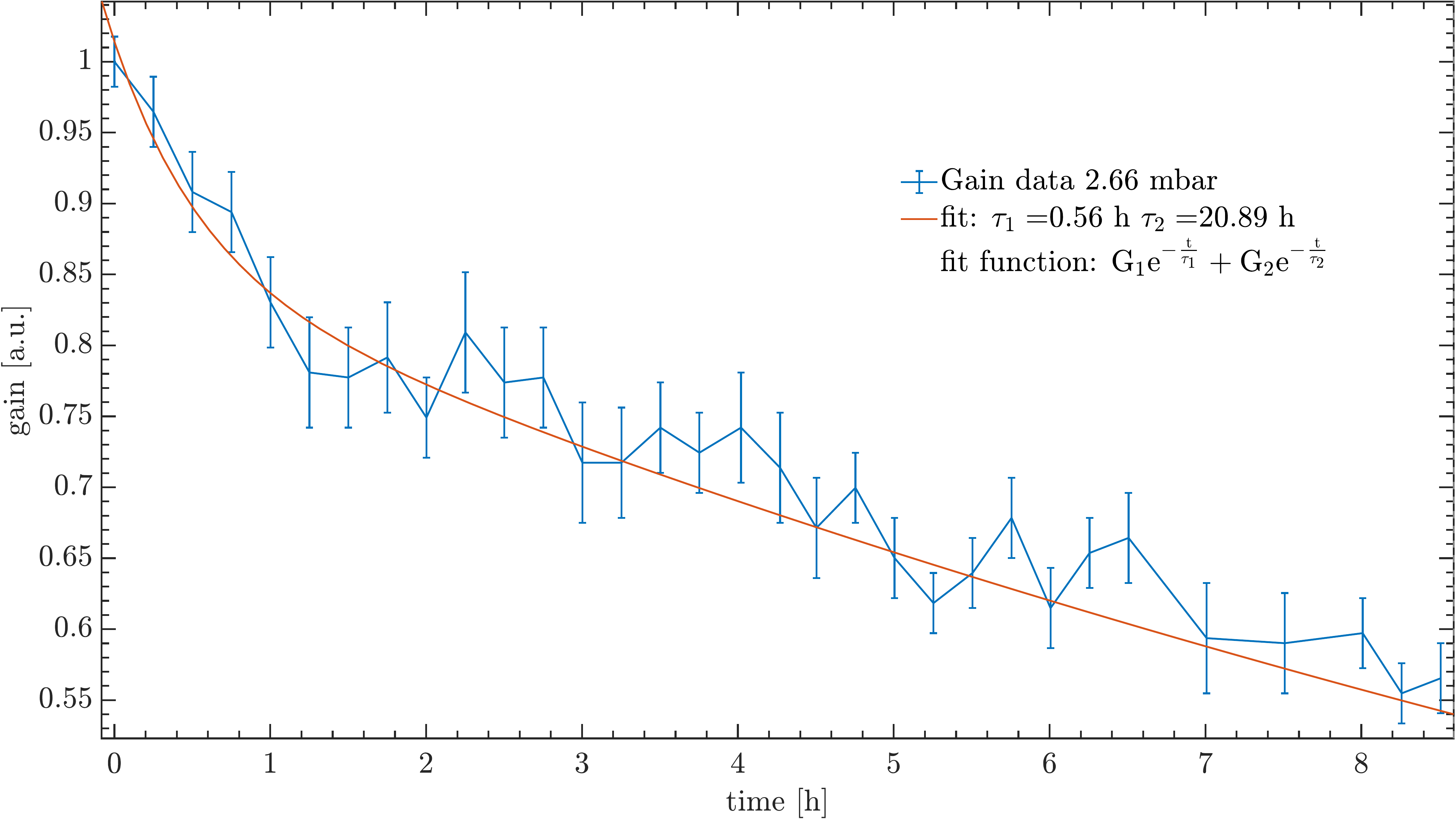}
	\caption{Example of gain decay vs time\label{fig_gain_time}}
\end{figure}
Fig \ref{fig_gain_time} shows an example of these measurements.
It easy to note a double exponential decay rate.
\begin{equation}
	\rm G(t)=G_1e^{-\frac{t}{\tau_1}}+G_2e^{-\frac{t}{\tau_2}}
	\label{eq_gain_time}
\end{equation}
Eq. \ref{eq_gain_time} represent the model function used to fit the gain vs time data.
Determining the parameters $\rm G_1$, $\rm G_2$, $\tau_1$, and $\tau_2$, we corrected the data gain behavior in function of the position along the glass tube diameter using Eq. \ref{eq_gain_time}.
Figs. \ref{fig_gain_1}, \ref{fig_gain_2}, and \ref{fig_gain_3} report the data after the time correction.
\section{Discussion}
\subsection{Data Comparison \label{subsec:data_compare}}
The main work with measurements about the gain at $\lambda=633$ nm in He-Ne induced by a RF plasma discharge has been made in 1973 by Tsarkov and Molchanov \cite{Tsarkov1973}.
But the differences in gas pressure and the ratio between He and Ne and also, the used discharge power make impossible a direct comparison with our results.\\
Another possibility could be a comparison with the work of Spoor and Latimer \cite{Spoor1984}.
In this case it is possible to find a couple of data-set similar in terms of gas characteristics and discharge power, but the DC supply system makes, again, the analogy impossible.\\
Nevertheless, we can find reasonable information that are still valid in our case: the order of magnitude of the gain, and the typical shape of the gain in function of the cavity radius.
About the first, it is easy to note that the values founded by Tsarkov and Molchanov, and Spoor and Latimer are of the same order of magnitude of the measurements shown in Figs. \ref{fig_gain_1},\ref{fig_gain_2},\ref{fig_gain_3}.
More specifically, the  data shown in Fig. \ref{fig_gain_1} are similar to the one of Fig. 3 of \cite{Spoor1984}.
If we convert the gain of our data into gain per unit of longitudinal length of plasma, we get values from $\rm 1 \% m^{-1}$ up to  $\rm 20 \% m^{-1}$, that is the same range of values shown in Fig. 3 of \cite{Spoor1984}.\\
It is also interesting to underline the presence of a peculiar behavior of gain in Fig. \ref{fig_gain_3}.
Near the center the gain curve of 1.33 mbar does not correctly follow the gain shape predicted by the model.
It is an effect of gain saturation that happens for very high RF power supply \cite{Spoor1984}.
\subsection{Ring Laser Gyroscope\label{subsec:RLG}}
Our main interest is to investigate the laser behaviour for RLG applications.
In particular to have a more detailed description of its active medium.\\
To correctly construe the data collected in this paper we have to remember that the evaluated gains values are referred to a single pass through the active medium, and that the optical power losses of our RLG are of the order of 100 ppm.
In this view an optimal working regime for a RLG should be the one with total pressure in the range between 5.00 and 10.00 mbar and output power RF generator of the order of few hundreds of mW.
The cited pressure range guaranties an homogeneous shape of the gain curve, and this level of RF power ensures a gain high enough to balance the losses, and, at the same time, keeping the system under the threshold of multimode operation.\\
We tested this range of pressure and RF power on our RLG GP2.
We tried several pressure level within the deduced range.
For each pressure level we tuned the supply current of RF generator in order to explore values, not only in the predicted range, but also higher and lower.
The conclusion is that in the pressure interval from 7 to 9 mbar and with a RF power of about 200 mW the RLG works in a transversal and longitudinal single-mode regime.
\section{Conclusion}
In literature the study on RF discharge systems have demonstrated that the efficiency of this supply method is better than the DC one \cite{Andrews1989}.
However, the results obtained in this work show that the transversal optical gain of a RF He-Ne plasma discharge has the same shape of that generated by means of a DC discharge.\\
Moreover, these results give us a series of relevant information to better understand the characteristics of a RLG.
To verify this information we have tested on our RLG GP2 a setup with the pressure and RF power in the previously indicated range of values (see Sec \ref{subsec:RLG}).
We obtained a monomode operation of RLG in the range between 7 to 9 mbar of He-Ne pressure and an output power of RF discharge system around 200 mW in accordance with the predicted values.


\begin{thebibliography}{10}
	
	\bibitem{Schreiber2013}
	K.~U. Schreiber and J.-P.~R. Wells.
	\newblock {Invited Review Article: Large ring lasers for rotation sensing}.
	\newblock {\em Review of Scientific Instruments}, 84(4):041101, 4 2013.
	
	\bibitem{Gebauer2020a}
	A.~Gebauer, M.~Tercjak, K.~U. Schreiber, et~al.
	\newblock {Reconstruction of the Instantaneous Earth Rotation Vector with
		Sub-Arcsecond Resolution Using a Large Scale Ring Laser Array}.
	\newblock {\em Physical Review Letters}, 125(3):033605, 7 2020.
	
	\bibitem{Capozziello2021}
	S.~Capozziello, C.~Altucci, F.~Bajardi, et~al.
	\newblock {Constraining theories of gravity by GINGER experiment}.
	\newblock {\em Eur. Phys. J. Plus}, 136:394, 2021.
	
	\bibitem{Belfi2014}
	J.~Belfi, N.~Beverini, D.~Cuccato, et~al.
	\newblock {Interferometric length metrology for the dimensional control of
		ultra-stable ring laser gyroscopes}.
	\newblock {\em Classical and Quantum Gravity}, 31(22):225003, 11 2014.
	
	\bibitem{Santagata2015}
	R.~Santagata, A.~Beghi, J.~Belfi, et~al.
	\newblock {Optimization of the geometrical stability in square ring laser
		gyroscopes}.
	\newblock {\em Classical and Quantum Gravity}, 32(5):055013, 3 2015.
	
	\bibitem{Beverini2019a}
	N.~Beverini, G.~Carelli, A.~Di~Virgilio, et~al.
	\newblock {Length measurement and stabilization of the diagonals of a square
		area laser gyroscope}.
	\newblock {\em Classical and Quantum Gravity}, 37(6):065025, 3 2020.
	
	\bibitem{Belfi2016}
	J.~Belfi, N.~Beverini, F.~Bosi, et~al.
	\newblock {First Results of GINGERino, a deep underground ringlaser}.
	\newblock {\em Review of Scientific Instruments}, 88(3):034502, 1 2016.
	
	\bibitem{Belfi2017}
	J.~Belfi, N.~Beverini, F.~Bosi, et~al.
	\newblock {Deep underground rotation measurements: GINGERino ring laser
		gyroscope in Gran Sasso}.
	\newblock {\em Review of Scientific Instruments}, 88(3):034502, 3 2017.
	
	\bibitem{Belfi:18}
	J.~Belfi, N.~Beverini, G.~Carelli, et~al.
	\newblock {Analysis of 90 day operation of the GINGERINO gyroscope}.
	\newblock {\em Appl. Opt.}, 57(28):5844--5851, 10 2018.
	
	\bibitem{DiVirgilio2020a}
	A.~D.~V. Di~Virgilio, A.~Basti, N.~Beverini, et~al.
	\newblock {Underground Sagnac gyroscope with sub-prad/s rotation rate
		sensitivity: Toward general relativity tests on Earth}.
	\newblock {\em Physical Review Research}, 2(3):032069, 9 2020.
	
	\bibitem{Mielenz1965}
	K.~D. Mielenz and K.~F. Nefflen.
	\newblock {Gas Mixtures and Pressures for Optimum Output Power of rf-Excited
		Helium-Neon Gas Lasers at 632.8 nm}.
	\newblock {\em Applied Optics}, 4(5):565, 5 1965.
	
	\bibitem{Mazanko1971}
	I.~Mazanko, N.-D. Molchanov, M.~Ogurok, and M.~Sviridov.
	\newblock {Gain distribution measurement in He-Ne laser tubes}.
	\newblock {\em Optics and Spectroscopy}, 30(495):927--931, 1971.
	
	\bibitem{Tsarkov1973}
	V.~Tsarkov and M.~Molchanov.
	\newblock {Measurement of gain distribution in the tube of a helium-neon laser
		($\lambda$=633 $\mu$m) uner high-frequency excitation}.
	\newblock {\em Optics and Spectroscopy}, 35(191):328--329, 1973.
	
	\bibitem{Spoor1984}
	S.~Spoor and I.~D. Latimer.
	\newblock {An accurate determination of the radial distribution of gain at 633
		nm in small bore helium-neon discharges}.
	\newblock {\em Journal of Physics D: Applied Physics}, 17(8):1607--1615, 8
	1984.
	
	\bibitem{Andrews1989}
	D.~A. Andrews and T.~A. King.
	\newblock {UHF excitation of helium-neon lasers: II. Comparison with DC}.
	\newblock {\em Journal of Physics D: Applied Physics}, 22(9):1315--1320, 9
	1989.
	
	\bibitem{Gray1996}
	B.~S. Gray, I.~D. Latimer, and S.~P. Spoor.
	\newblock {Gain measurements at 543 nm in helium neon laser discharges}.
	\newblock {\em Journal of Physics D: Applied Physics}, 29(1):50--56, 1996.
	
	\bibitem{Bilger1995}
	H.~R. Bilger, G.~E. Stedman, Z.~Li, U.~Schreiber, and M.~Schneider.
	\newblock {Ring Lasers for Geodesy}.
	\newblock {\em IEEE Transactions on Instrumentation and Measurement},
	44(2):468--470, 1995.
	
	\bibitem{Graham2006}
	R.~Graham.
	\newblock {\em {Ring Laser Gain Media}}.
	\newblock PhD thesis, University of Canterbury, 2006.
	
	\bibitem{McLeod2000a}
	D.~P. McLeod.
	\newblock {\em {Seismic effects in ring lasers and transverse mode selection in
			helium-neon lasers}}.
	\newblock PhD thesis, University of Canterbury, 2000.
	
\end{thebibliography}

\end{document}